# REYSA BERNSON, THE UNCONVENTIONAL HEAD OF THE FIRST FRENCH PLANETARIUM

**Yaël Nazé**

*University of Liège, B5C, Allée du 6 Août 19c, B4000-Liège, Belgium.*
E-mail: ynaze@uliege.be

**Abstract:** The first modern planetarium was presented in 1923 in Jena, Germany. Very soon in the subsequent years, planetariums were installed in other parts of Europe as well as in America. France, however, got its first planetarium only in 1937, for the World Exhibition organized in Paris. The team that took care of that planetarium was headed by a female amateur astronomer named Reysa Bernson. This choice might seem surprising, but it was not made at random, thanks to her never-ending astronomical activities at that time. This paper aims to bring back memories of this very active amateur astronomer of the 1920s and 1930s, and show the many ways in which astronomy was disseminated a century ago.

## 1 INTRODUCTION

Public outreach has a long history in astronomy, with a widening over the years of the types and size of the targeted public as formal education became more generalized in society. Several outreach means were of course explored, the most common being writing (books, articles) and talks. In this field, the skills of several people have been particularly recognized over the centuries, notably Agnes Clerke (1841–1907) and Camille Flammarion (1842–1925), to take two major figures from the second half of the nineteenth century. After World War I, the dissemination efforts increased. More amateur astronomers became involved, as this activity was no longer reserved for a few wealthy individuals, as in the previous century. More articles appeared in more numerous newspapers and magazines. More outreach venues were also developed, including science museums and planetaria (whose first example was built in Jena, Germany in 1923).

In this context, a prominent figure emerged in France, and her name was Reysa Bernson. She was active in several domains, including amateur observational astronomy, presentations at schools, and talks and articles for the public at large. While she was a prominent outreach figure at the time, her contributions are almost totally unknown today. In this paper, after discussing her early years I will examine her different types of astronomical activities, and highlight her numerous achievements. The concluding section will then look into the recognition that she received and the obstacles that she had to face.

## 2 THE EARLY YEARS

Born in 1904[1] in Lille, a large city in northernmost France, Reysa Bernson (Figure 1) was the only child of two MDs, Désiré Verhaeghe (1874–1925) and Dweira Bernson (1871–1944). Her father was a French citizen from a middle-class family from the north of France, while her mother was a refugee from Brest (Belarus) who had fled the per-secution of Jewish people following the death of Tsar Alexander II (Delmaire and Faidit, 2017). They married in Belgium in 1900, just after her MD thesis defense but a few years before his. Their political views clearly leaned to the left of the spectrum. Indeed, the two of them were deeply involved in the efforts to improve the health situation of local workers, through writing articles that supported improved health care or their work in dispensaries (for tuberculosis patients for him, and for child-and-mother care for her). Furthermore, Désiré Verhaeghe was a member of the French Workers' Party and later of the SFIO (French section of the Workers' International). He later became the counsellor of the socialist mayors of Lille, where the family lived.

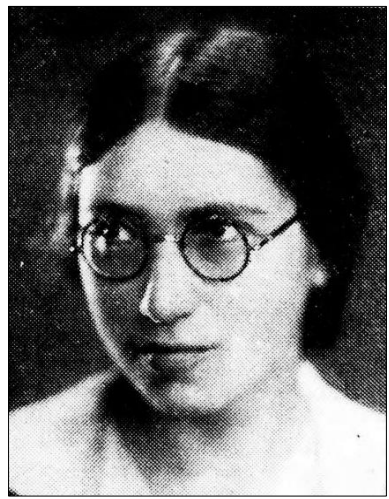

Figure 1: Reysa Bernson (1904–1944) in 1932 (after M.D., 1932; courtesy: Gallica).

However, the marriage did not last. Just after World War I, Désiré Verhaeghe left the marital home and moved in with another woman (with whom he had two children, boys born in 1921 and 1925). Désiré and Dweira never officially divorced, and the split left deep scars. Dweira Bernson therefore later opposed any memorial to her husband (Anonymous 1927; 1928), although a local school finally was named after him, in 1931. His daughter, Reysa (1904–1944), even came to the inaugural ceremony and gave an astronomical talk there (see Bernson, 1934). As a consequence, starting in 1920, Reysa only used her mother's maiden name, and was known as Reysa Bernson.[2]

Because she obtained multiplied diplomas, Bernson remained a student for quite some time (Barré-Lemaire, 2019; see also local newspapers listing graduates each year). She obtained the two parts of the 'baccalauréat', the French exam at the end of secondary school, first in 1921 for sciences and literature and then in 1922 for mathematics. While finishing her secondary studies, she joined Lille University to study Russian, which led in 1923 to her first diploma (in 'Études Russes', i.e. Russian studies). The following year, she graduated as 'Licenciée ès Sciences', more or less equivalent to a modern Bachelor's degree, and obtained a 'Certificat d'Études Supérieures' in Astronomy. In subsequent years she obtained additional certificates in general chemistry and radio-telegraphy. Finally, she secured her 'Diplôme d'Études Supérieures' in 1927, more or less equivalent to a modern Masters degree, and a diploma in 'Astronomie Approfondie' (advanced astronomy) in 1934, both again from Lille University.

During her studies, she was actively involved in student representative bodies (Barré-Lemaire, 2019; see, also, newspapers of the time). From 1925 she belonged to the Organising Committee of the Lille Association of Female Students and was the Head Editor of its magazine, called *Lille-étudiante*. In the following year, she was an official Lille delegate at the Annual Meeting of the UNEF (National Union of French students) as well as a UNEF delegate in Geneva at the Congress of the International University Federation for the League of Nations, the ancestor of the United Nations. In 1927, she became Treasurer of the Lille association of all (male + female) students and their officer in charge of external relations. She was elected the same year as Vice-President of the UNEF, a position in which she remained in 1928. In addition, she was one of the student representatives on the Organizing Committee of CTI (Intellectual Workers Confederation) from 1929 to 1934. Finally, until 1932, she was also in charge of the Career Guidance Office, which was created within UNEF in 1927 to help students cope with the severe employment problems faced by new graduates. This implied collecting information on which job could be reached by which diploma and assessing the bottleneck status of these jobs, to allow students to choose studies leading to more secure employment possibilities after graduation. Unfortunately, there was one unpleasant incident relating to this Directorship. In 1930, at the annual UNEF meeting organized in Alger, the Lille delegation decided not to back her candidacy for the Directorship of that office (Anonymous, 1930). However, she got immediate support from the student delegation from Strasbourg University (Anonymous, 1930). She then won the election that year, and in subsequent ones. She resigned a few years later, following rumors that she was "… working against the students … [and] favoring female students …", which she denied (Fred, 1933; see Barré-Lemaire, 2019).

Testimonies of Bernson's strong involvement in career guidance can be found in the articles she wrote (Bernson 1927; 1929a; 1931a); the talks she gave on the subject, including one "feminist tea" in May 1930; and the media interviews she gave. She is quoted as an authority on the the subject in numerous French newspaper articles between 1927 and 1933. However, the Lille student delegation of 1930 was not the only one showing prejudice towards her, as is obvious from the following newspaper text:

> Mums, what are you going to do with your daughters? I went to ask Miss Reysa Bernson, who may never be a mother because she only cares about career guidance, physics and astronomy. Her flat, at 219c boulevard de la Liberté in Lille, is a pure curiosity that reveals the originality of its occupant: black furniture with baroque geometric lines, overly subdued lighting that covers two thirds of the study in shadow, books everywhere, a map of the Moon on the back of a cupboard door, all in a setting of sober, fine elegance. The lips, cheeks and eyelashes of this stern-faced 'miss', with her tortoiseshell glasses and long hair in bands and braids, have never been touched by the slightest grooming device. (Fred, 1933).

This same article concludes that a wedding is the best career for female students, which may explain some of the previous text.

It is not known what exactly triggered Bernson's interest in astronomy. In particular, there was no sign of any interest in that domain by her parents or by close relative or friends. There is, however, some information about her childhood which gives hint of early influences. A newspaper article following her Henri Rey prize (MD, 1932) mentions her interest in the hybrid solar eclipse of April 19[3] and her surprise at discovering that it wouldn't be a holiday to allow all students to observe it. At the same time, her parents bought her a subscription to a children journal called *Les Petits Bonshommes* (Mathieu, 2022), in which Jean Couture, a former member of the SAF (Astronomical Society of France), held a regular astronomical chronicle. Finally, she mentioned receiving as a gift famous popular books by Camille Flammarion, whose contents she presented to her school friends afterwards (Bernson, 1933a). In any case, she became a SAF member in 1920 (Anonymous, 1920), at the age of only sixteen! She would become a perpetual SAF member eight years later. It is after this affiliation date that her astronomical activities start to be documented. The following sections aim at presenting each facet in turn.

### 3   AMATEUR ASTRONOMER ACTIVITIES

There are two clear parts to her amateur astronomer activities: on the one hand, she participated to the organizational aspects and, on the other hand, she was a keen observer of the sky.

#### 3.1   Organizing

Bernson participated in various SAF activities (articles, talks, celebrations) from 1920 until 1940. The last report is her participation in the 15[th] anniversary of Flammarion's death in his property of Juvisy, in the outskirts of Paris, in June 1940 (Anonymous, 1940). However, she always remained a 'simple' SAF member, never nominated to any official position.

In contrast, she headed the development of an astronomical society in northernmost France. In fact, after participating to a General Assembly of the SAF in December 1922, she realized that the Parisian amateurs had access to a lot of facilities: a common observatory, a common library, regular courses and talks (Bernson, 1933a). However, for people outside the capital city, the situation was often quite different. In particular, in the north of France amateurs were scattered and isolated, hence they hardly ever met to share information or instruments. She therefore asked the SAF librarian for a list of members living in Lille and neighbouring cities. She then went and contacted each of them individually, asking whether a local amateur society would interest them. A first face-to-face meeting occurred in April 1923 with twelve 'good will' people (including her) paving the way for the founding of a new society, the AAN (Astronomical Association of the North). She then obtained from the University authorities access to a room in the Physics Institute for lectures and to the terrace of the same building for observations. The grand opening officially took place in mid-December 1923 with a public talk and, two months later, the AAN already boasted 150 members.

Bernson never chaired the AAN. From the start, she was only the General Secretary and she remained so until the General Assembly of January 1938 (Anonymous, 1938). However, as the *AAN Bulletins* show, she was a clear pillar of the Association, being its founder, writing the majority of the text in the *Bulletins*, being a contact point for media as well as for other amateurs, giving a large number of the AAN talks, and engaging in all of its school activities (see the next section). A sudden stop of her activities occurred in 1936, most probably linked to the disastrous eclipse expedition (see below). It marked the end of her involvement. The Adjoint Secretary, Henri d'Halluin, then took over her duties. From then on, her name never appeared again in the *AAN Bulletins* of the 1930s (e.g. her subsequent prizes, or the 1937 planetarium she was directing, were never mentioned), nor was she invited to give another talk. In addition, she never became an Honorary President or even received a 'Thank-you" letter. One day she simply disappeared from the very association that she had founded. Post-war anniversary issues only mentioned her name quickly, as a founder, but without any details.

In parallel, as an active amateur, Bernson realized that some amateurs (herself included) aspired to perform true astronomical research. Here the modest size of amateur instruments was compensated for by long stretches of observing time, which could provide useful results (as indeed happens nowadays with pro-am collaborations). However, Bernson (1931b) identified a major organizational problem: data storage and analysis. If a local astronomical society published a journal, at best it could print the measurements made by local amateurs, but these journals often stayed on bookshelves accumulating dust. Then even the most motivated amateurs could become discouraged and stop observing. At that time, professional astronomers had started to organize themselves at the international level, with the founding of the International Astronomical Union (IAU) in 1919 (Sterken et al., 2019). For amateurs, no such thing existed as there were only local or national groups, generally focused on a specific subject (for example, in France, the various 'Commissions' of the SAF, or the AFOEV, the French Association of Variable Star Observers, both of which she participated in—see Section 3.3, below).

Bernson therefore thought of creating another international organization, this time devoted to amateur astronomy. It may be interesting to note here that she had already put forward similar internationalist views and insisted on cooperation during her involvement in student affairs. Here, the idea was to centralize project requests from professional astronomers or 'enlightened' amateurs (i.e. any astronomer able to analyze physical phenomena). Such projects would have a specific goal and a clear methodology, which an amateur could simply follow with an instrument—this would also help standardize the work amongst different amateurs, facilitating the interpretation of the collected data. Then the amateur observations would be collected in a single place, to facilitate their diffusion. There would only be a small risk for data to never be used, since they were requested by someone in the first place. From 1924, she made this proposal at first informally, talking of the idea to amateur colleagues as well as observatory directors whom she was meeting.

Then she decided to go public, through a talk at the annual meeting of the French Association for the Advancement of Science in Alger (Bernson, 1930: 59) and an article published in both the *AAN Bulletin* and *L'Astronomie* (Bernson, 1931b). The proposal was further presented by Mrs Gabrielle Flammarion at the meeting of French astronomers organized in Paris in July 1931 (which she attended).[4] Despite the initially warm reception, nothing was done in practice to set up such an organization and her idea was simply forgotten. It must be said that localized efforts continued afterwards (e.g. see the websites of the AFOEV, AAVSO, BeSS, etc. and the numerous publications associated with them). Currently (2023), the IAU is actively developing pro-am relations and projects. Maybe her proposal simply came too early.

As soon as she became active in astronomy, Bernson saw it as a duty to be a 'transmission wheel'—one would use the word 'networking' nowadays. It can be suggested that she already had good experience in that domain thanks to her previous activities in student organizations. In fact, each time she was going to some place, whatever the reason (e.g. UNEF activities), she tried to meet local amateur as well as local professional astronomers, and such endeavours are particularly documented during her trips to Germany in 1931 and Poland in 1932. In turn, they sent her their observations which she duly transmitted to the *AAN Bulletin* or the *SAF Journal*. While this activity was important and useful for both sides, she rarely was acknowledged in cases where there were positive outcomes. For example, when in Potsdam in 1931 she met Robert Henseling, the head of the important German amateur astronomical association, the 'Bund der Sternfreunde'. She wrote she was stunned

> … to learn that the Bund der Sternfreunde is only in correspondence with an isolated amateur living in the south of France. (Bernson 1931c).

She immediately added the AAN as an additional contact, and worked to make other French additions afterwards. Five years later, the SAF eclipse expedition passed through the Berlin area and was received by this now well-known acquaintance. Henseling was friendly, and his "… desire to please us was apparent in the smallest details …" (Bidault de l'Isle, 1937), but the only identified correspondent in that publication is Henri d'Halluin which would soon replace her as AAN Secretary.

A final fact may be added in this organizational sub-section. In 1938, Armand Delsemme (1918–2017), was then a young university student who subsequently went to have a career dedicated to cometary studies, and he decided to create an astronomical society in the Belgian city of Liège where he studied. The previous year, he had been part of Bernson's planetarium team and she had impressed him very much (Delsemme, 1989). He therefore asked her to give the inaugural talk and to be the 'godmother' of the new society, both of which she accepted (see the transcript of her talk in Bernson, 1939a). Contrary to the *AAN Bulletins* after 1937, her name is mentioned several times over the years in *Le Ciel*, the magazine of the SAL (Liège Astronomical Society), especially in anniversary issues in which her planetarium activities and mentorship are always underlined, along with the honor she bestowed on Delsemme by accepting him on her team, and finally by accepting the SAL patronage. These all demonstrate the great appreciation SAL members had for her.

### 3.2   Observations

There is abundant evidence of Bernson's observing activities. One useful source is *l'Astronomie*, the SAF magazine. This journal regularly reported observations by SAF members of var-

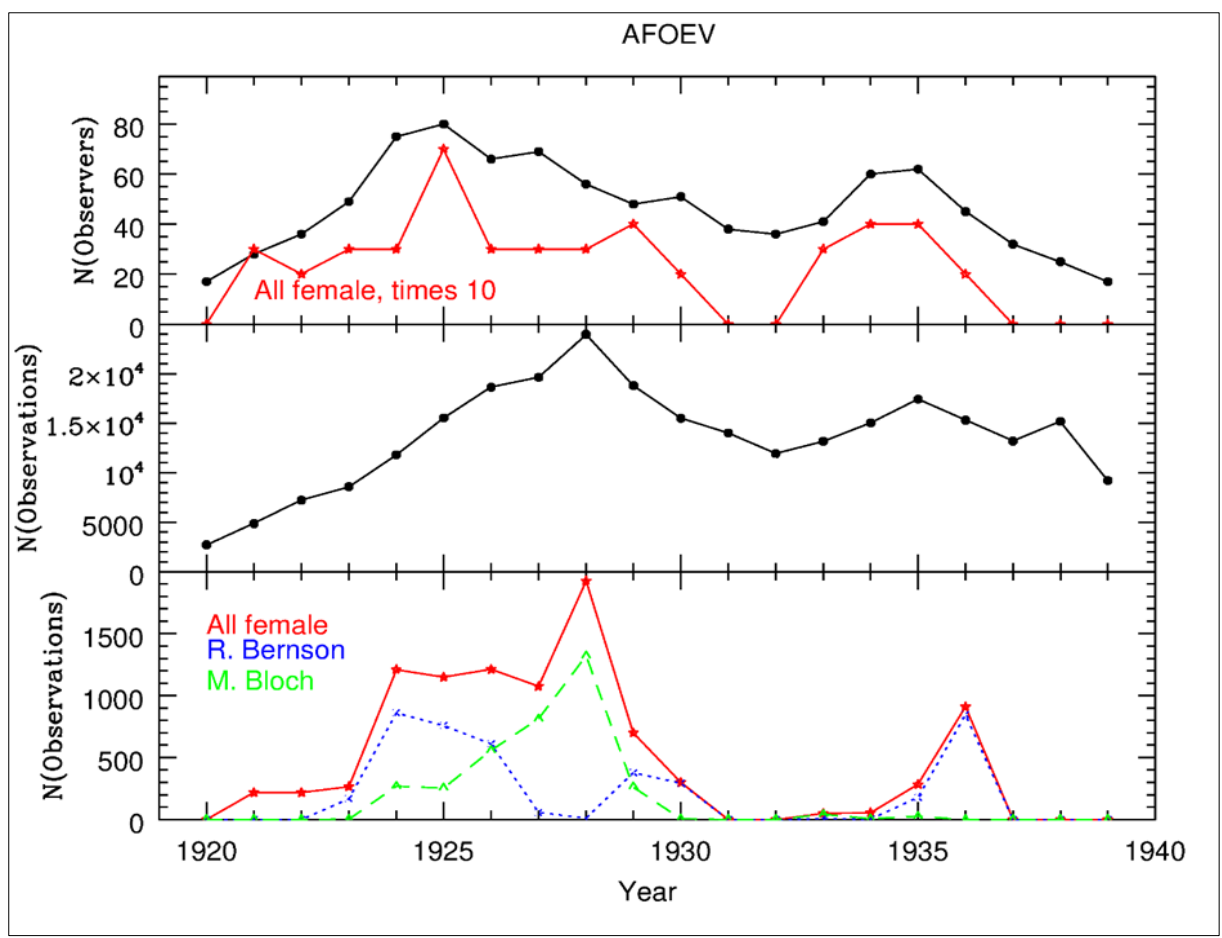

Figure 2 (top to bottom): Evolution with time of the number of active observers (i.e. persons with at least one observation that year) with the number of female observers in red, and evolution with time of the number of observations for all observers (middle panel) and female observers (bottom panel, in red). The contribution of Reysa Bernson (BNN) is shown with a dotted blue line and that of Marie Bloch (BLO) with a green dashed line. (plot: Yaël Nazé).

ious phenomena, and with varying levels of detail. The issues from 1921, 1924, 1925, 1926, 1928, 1930, 1931, 1932, 1933, 1935, 1936, and 1939 all mention her name in such observing reports. The majority of these reports are quite short, as for other SAF members, with only a few providing actual specifics (e.g. the time of first contact during an eclipse). Nevertheless, these reports reveal that she had a broad interest in all celestial phenomena observable at the time: meteors, sunspots, solar and lunar eclipses, transits of Mercury, and lunar occultations.

Of much higher quality was her work for AFOEV, whose archives are available online (https://cdsarc.u-strasbg.fr/afoev/). One can count 229 observers who enrolled with the AFOEV between 1900 and 1940. Amongst them, 14 are identified as female (i.e. 'Miss', 'Mrs', or with a female surname). Reysa Bernson appears with an enrollment date of 1923, and her official acronym in the database is BNN. Figure 2 graphically illustrates the activity of these French amateur observers between 1920 and 1940. As can be seen, the overall evolution with time of the number of active observers does not depend much on gender, but the number of observations is however gender-dependent, with a lack of data by female observers in the early 1930s. Another point to note is the large contribution of BNN compared to other fellow-female observers. In fact, at the time, only two female amateurs made an important contribution: BLO (Marie Bloch) and BNN (Reysa Bernson). Together, except for one year, they account for at least two thirds of the reported data, and often more than 90%. Bloch was an assistant at Lyon Observatory (first unpaid, then with a salary). One then would expect to find non-negligible observing activity from her as it was part of her duties. Such was not the case of Bernson, which renders the quantity of her contributions even more remarkable. Her 4171 data points (Figure 3) correspond to 62 different objects which can be categorized, using the Simbad

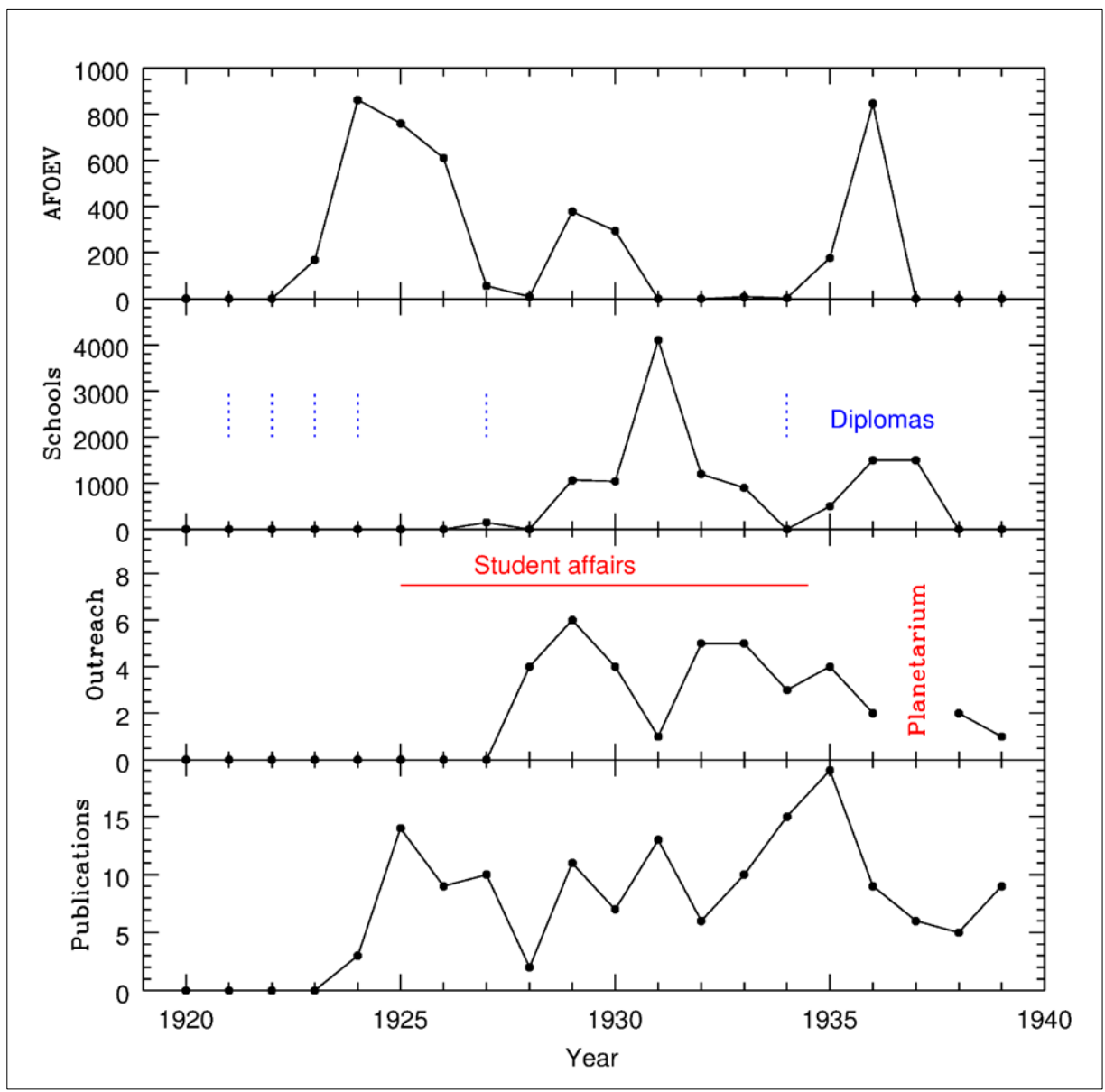

Figure 3 (top to bottom): Evolution with time of Reysa Bernson's activities (variable star observations recorded at AFOEV, number of secondary school students assisting to her talks, number of outreach talks, number of publications). No point is provided for outreach talks in 1937 as the huge number of planetarium shows at the Paris Exhibition cannot be compared to the typical number of talks in other years. For school students in 1936–1937, only the global number is provided in Bernson (1937b), not the distribution between the two years: here we assume an even splitting. Note that the information on talks and publications may be incomplete as only those available in public archives are counted (*AAN Bulletins* and public databases, and the BNF, the National French Library, Gallica, and from the 'mediathèque de Roubaix'). Publications reproduced in another journal are not counted separately. A biographic timeline as well as her publication list are available from the U. Liège repository, https://orbi.uliege.be/, as a complement to this paper (plot: Yaël Nazé).

database, as Mira variables (32), long-period variables (23), novae (2), and other variable stars (5). Considering the cloudy skies of northern France, it was logical for her to focus on long-term variables as short-term magnitude changes could easily have been missed because of bad weather.

One major event caught Berson's attention: Nova Herculis 1934. As with many amateurs at this time, she followed the evolution of this nova with great interest. A major puzzle at the time was a secondary brightening after an initial luminosity decrease. In Bernson (1935), she explained that she drew the light curves of several novae not by considering the usual apparent magnitude ordinate but rather by assuming that all novae reached the same peak luminosity. This led her to propose two families of novae: a large group showing light curves with an exponential decrease and another group, including Novae Aurigae 1892 and Herculis 1934, showing light curves with a shallow increase before the peak and then a long plateau. She added that some novae, whatever their group classification, could present rapid fluctuations. Apparently, she actively promoted this dichotomous model among French amateur astronomers (see Grouiller, 1936), but publishing in French in an amateur astronomy journal did not help make her ideas known beyond the SAF circle. Meanwhile, professional astronomers develop their own classifications of novae at about the same time. Lundmark (1935) identified five different types of novae based on their light-curves: flash, flash-oscillation, wave, wave-oscillation, and jump novae. Gerasimovic (1936) came up with just two types, slow and flashing novae, while Mc Laughlin (1939) also suggested two types of novae, slow and fast novae. It is significant that all of these schemes (including Bernson's) have similar features.

Bernson was never employed by a professional observatory, and in retrospect one may wonder why. After all, she had adequate academic qualifications and also extensive observing experience, at the time when a number of women worked in French observatories, with or without salaries. For example, the proceedings of the 1935 IAU General Assembly in Paris lists 21 female participants, 12 being part of the French delegation. They represented 13% of the French delegation, which is a remarkable number considering the epoch (e.g., the fractions for the USA and UK delegations were 8% and 5%, respectively). It would seem that Bernson could have made professional astronomy her career, had she wished to, and this may even have saved her life.[5]

That said, Bernson did participate in at least one official professional astronomical research project: the Bureau des Longitudes 1936 solar eclipse expedition (Bernson, 1939b). For this, she received help from the Dean of Lille University, and funding from the CTI. Her project was to take infrared photographs of the corona. She had a special photographic camera built for this purpose, and she was allowed by Daniel Chalonge (another participant in this expedition) to train in its use at the photographic laboratory of the Paris Observatory. Unfortunately, bad weather affected the expedition, so no usable results were obtained. This must have affected her deeply, as she took sick leave from her AAN duties afterwards (Bernson, 1936)—it may be noted that the publication of this account was delayed, perhaps on purpose). In this context, it is important to note that when the expedition reports, including the one in *L'Astronomie* (Bureau des Longitudes, 1937), were published, no mentioned at all was made of Bernson's project or of her participation in the expedition. This surprising omission probably explains why she 'set the record straight' by publishing her own account two years later (Bernson, 1939b).

## 4  PUBLIC OUTREACH ACTIVITIES

In the 1920s and 1930s, Bernson was very active in the field of public outreach (see Figure 3). Her activity, while not reaching the exceptional level of Camille Flammarion, can only be compared to that of the most active popularizers of the time, e.g. Lucien Rudaux. Her activities were varied, basically covering all outreach possibilities except writing books (Figure 3). For example, she wrote more than one hundred outreach articles, mostly in the *AAN Bulletin* and in *L'Astronomie*. Outside of the planetarium animations, there are reports of ~40 outreach talks to amateur societies (e.g. the AAN, SAF, SAL, Astronomical Society of Picardie) as well as in popular venues (the Université Populaire) or high-society ones (e.g. the Société Dunkerquoise pour l'Encouragement des Sciences, des Lettres et des Arts, in Dunkirk).

For both talks and articles, three categories of themes can be found: general presentations (e.g. content and scale of the Solar System, the Milky Way, or the Universe), news (e.g. Pluto's discovery, Nova Herculis, contemporary US and Soviet stratospheric expeditions), and personal memories (e.g. travels to Germany or Poland, or the foundation of the AAN). From the articles and conference transcriptions, her style appears very appealing, with many comparisons linked to everyday life, anecdotes, a lively prose, and a touch of humor. She was keen to illustrate her talks with a lot of photographs, and even movies. Finally, she never hesitated to answer media requests, and even wrote in some newspapers herself (e.g. the unexpected meteors of 1933—Bernson, 1933b). From 1929, one can also find her talking on radio, e.g. for a live session for the eclipse of 2 April 1931 or for an astronomical series on PTT-Nord[6] the following years.

From feedbacks at the time, it seems clear that her contributions were much appreciated:

> Miss Reysa Bernson came last Friday to give, under the auspices of the Dunkirk Society, the high-level conference that it is customary to hear in this Company. The rooms of the town hall also welcomed a large and attentive audience who had come for the stellar journey promised by the speaker. It was a wonderful trip and Mr. Terquem, presi-dent of the Society, was right to congratulate himself on having chosen such a fine interpreter to open our winter lecture series … Miss Reysa Bernson's talk had the order and precision of a scientific insider. She kept it simple, even in the most complex areas of astronomy … Leaving the stellar domain, Miss Bernson knew how to captivate her audience by filling us with new and delicious details about our sister planets and our closest neighbor: the Moon. And it was for her an opportunity to set the record straight on certain scientific prejudices concerning them. Add to this a reliable interpretation of a whole collection of unique photographs and you will have a very imperfect idea of the charm of this lecture … and much more were presented, discussed and made, what's more important, friendly and understandable. (Anonymous, 1929).

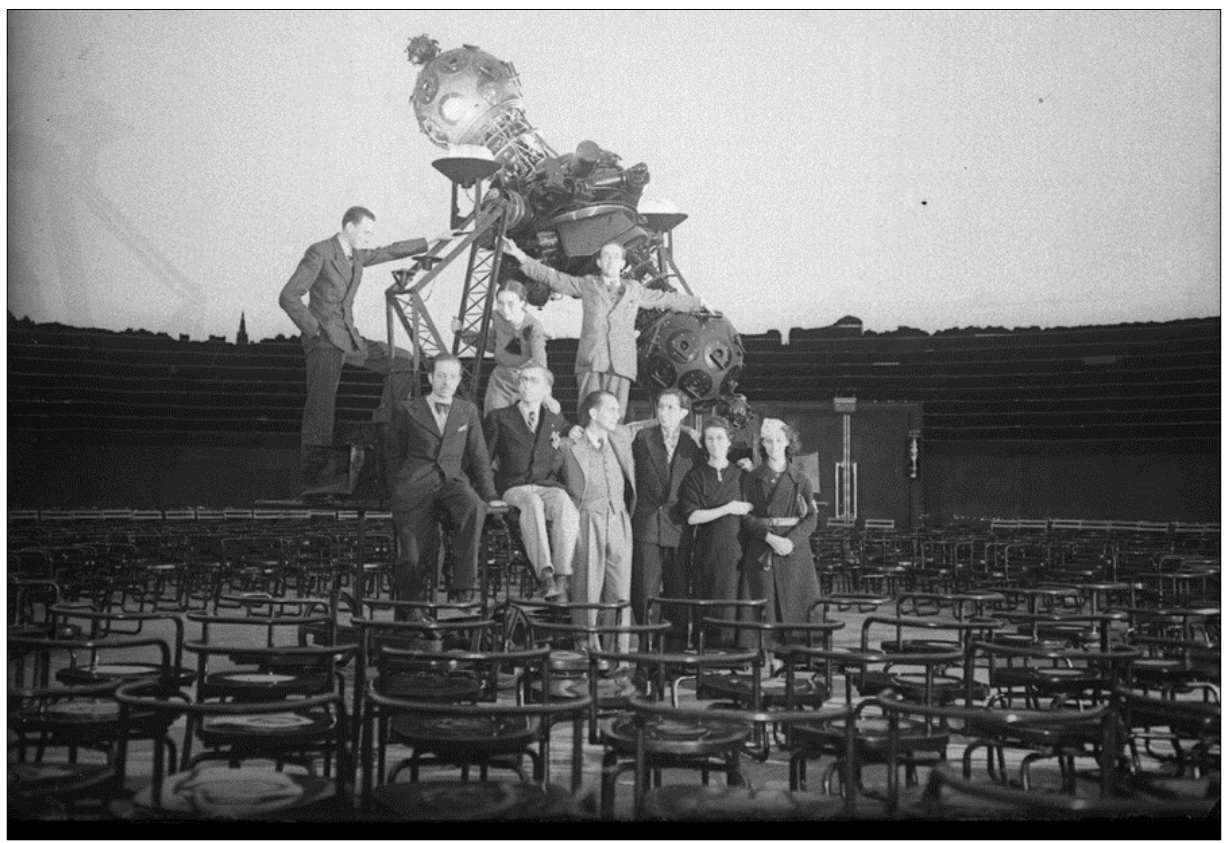

Figure 4: The 1937 planetarium team (minus one), plus two young women (probably usherettes). Reysa Bernson is close to the projector, with André Hamon and Armand Delsemme in front of her (courtesy: Archives de l'Association Astronomique du Nord; © Association Jonckheere).

During her trip to Germany, Bernson (1931c) attended a planetarium session in Dresden. She immediately saw the outreach potential of such a facility, so she took every opportunity to praise planetariums while recalling their absence in France (notably while mentioning new installations in Netherlands or Belgium). It may thus appear unsurprising that she was chosen to head the first French planetarium, built for the 1937 World Exhibition in Paris (Figure 4). However, we must recall that this occurred about 90 years ago and offering such a position to a non-professional astronomer, and what is more to a woman, was highly unusual.[7]

Bernson's full planetarium team comprised the SAF Secretary André Hamon, the self-taught astronomer Eliezer Fournier, the astro-photographer Henri Kanapell, the observers Jacques Codry and Auguste Budry, as well as two young students, Armand Delsemme and Gérard Oriano.[8] In this context we may note that André Hamon (1900–1986), was Bernson's senior by four years and was SAF Secretary, yet he was her subordinate at the planetarium. Thus, choosing her for the task was far from obvious but, unfortunately, there is no public report of the exact process that led to it. We may speculate that maybe the choice of Bernson was linked to the nomination of André Léveillé (1880–1963) as General Secretary of 'Group I – Thought Expression' for the 1937 Exhibition. He was not only from Lille but also was Vice-President of the CTI in 1925–1939 (Bergeron and Bigg, 2015), which makes two possible direct links to Bernson.

The 1937 Exhibition actually included three parts devoted to astronomy. The best-known one was of course the astronomical exhibition of the 'Palais de la Découverte', supervised by Paris Observatory (see Figure 5). The Palais was specifically built for the exhibition, to showcase scientific achievements in an active way, and it has sur-vived to this day with the same spirit. A meeting report shows that the idea of including a plan-etarium within the Palais was discussed origin-ally but was soon discarded because of the cost and the fact that the "… audience tire of the show fairly quickly." (Anonymous, 1935). The other two astronomical parts were thus not integrated within the prestigious new Palais, but rather relegated to the 'Concessions', a sort of amusement park with a science flavor installed along the quayside. One was the Stellarium, a mock rocket journey through the Universe. This was a private initiative, without detailed scien-tific animations. The other initiative was the planetarium. Despite a lack of interest by the Palais builders, the planetarium was was soon approved for the Exhibition. There was already a mention of installing a planetarium in Paris in a 1934 letter from the French Ambassador in Berlin to the Director of the Luxembourg Mus-eum in Paris (Bergeron and Bigg, 2015). The successes of the planetariums at the Chicago and the Brussels World Exhibitions were noted by the French organizers of the Paris Exhibition and also by the Zeiss Company. As a conse-quence, a contract was signed and, starting in Fall 1935, articles in French newspapers about the future Exhibition mentioned the inclusion of a planetarium, often presenting it as a highlight, or a 'must-do'.

The planetarium team is seldom mentioned in newspaper articles of the time, which rather focused on the living experience of a planetar-ium session. Nevertheless, a few of them did so, and Bernson then was clearly identified not only as the planetarium head but also as a highly competent person:

> This is one of the highlights of the exhibition. It will be presented by one of our most learned astronomers, Miss R. Bernson. (Thoraval, 1937).

Or, the planetarium show was "… presented to us by a distinguished French astronomer, Mlle R. Bernson." (Anonymous, 1937a). More de-tailed praise can also be found in these three articles:

> Before, with Miss Bernson and the swarm of young astronomers surround-ing her, chatting about the joysticks and the multiple levers, I said: what tech-nique! Afterwards, looking at the trans-figured faces of the adults and children applauding and rising to their feet, my-self still spellbound for a long time, I thought this time: no, this is art. (Bour-del, 1937).

> The president listened attentively to the perfect explanations given by Miss Bernson. (Morice, 1937).

> We sail like the good old Cyrano de Bergerac, through the stellar spaces, and we would not be surprised, later on, when we return to Earth, to have scorched ourselves too much by rub-bing a star too closely. We sail, but at the whim of a voice. Filled with grat-itude for this voice, its spirit, the clarity of its presentations, the finesse of its humor, I bravely assumed that it was attached to a body. My questions were answered with the awful word 'Madame speaker?' and I was led to a young girl, Miss Bernson, of absolutely astral sim-plicity, who does not seem to suspect that she has an incredible talent, and who assures me that her other col-leagues do just as much and just as well. We were therefore guided by a voice, which was for us immaterial and wise, the voice of light itself, or of some archangel responsible for dusting the astral system. (Manceron, 1937).

A search amongst newspapers does not reveal any negative report relating to the planetarium or its team.

When the exhibition closed after five months, between 350,000 and 500,000 visitors are estimated to have come to the planetarium (the numbers reported in the newspapers var-ied enormously; e.g., see Albert, 1937; Anony-mous 1937b; 1937c). This huge success did not lead, however, to a permanent exhibition as at the Palais. In fact, the planetarium was simply dismantled, to the dismay of the media, and then sold in November 1938 to the city of Paris. It was then stored in the cellars of the National Conservatory of Arts and Crafts, finally to be dusted off and installed in 1952 … in the Palais, which originally did not want it!

For the SAF, the position with respect to the planetarium has been ambiguous: on the one hand, not a word is said about it in the 1937 volume of *L'Astronomie* while the Palais' as-tronomical exhibition is mentioned in detail and praised; on the other hand, the planetarium team won the commemorative medal in 1938 as the team recruited many new members for the Society thanks to their planetarium shows and related activities (talks, observing sessions) that they organized for exhibition visitors on the SAF grounds.

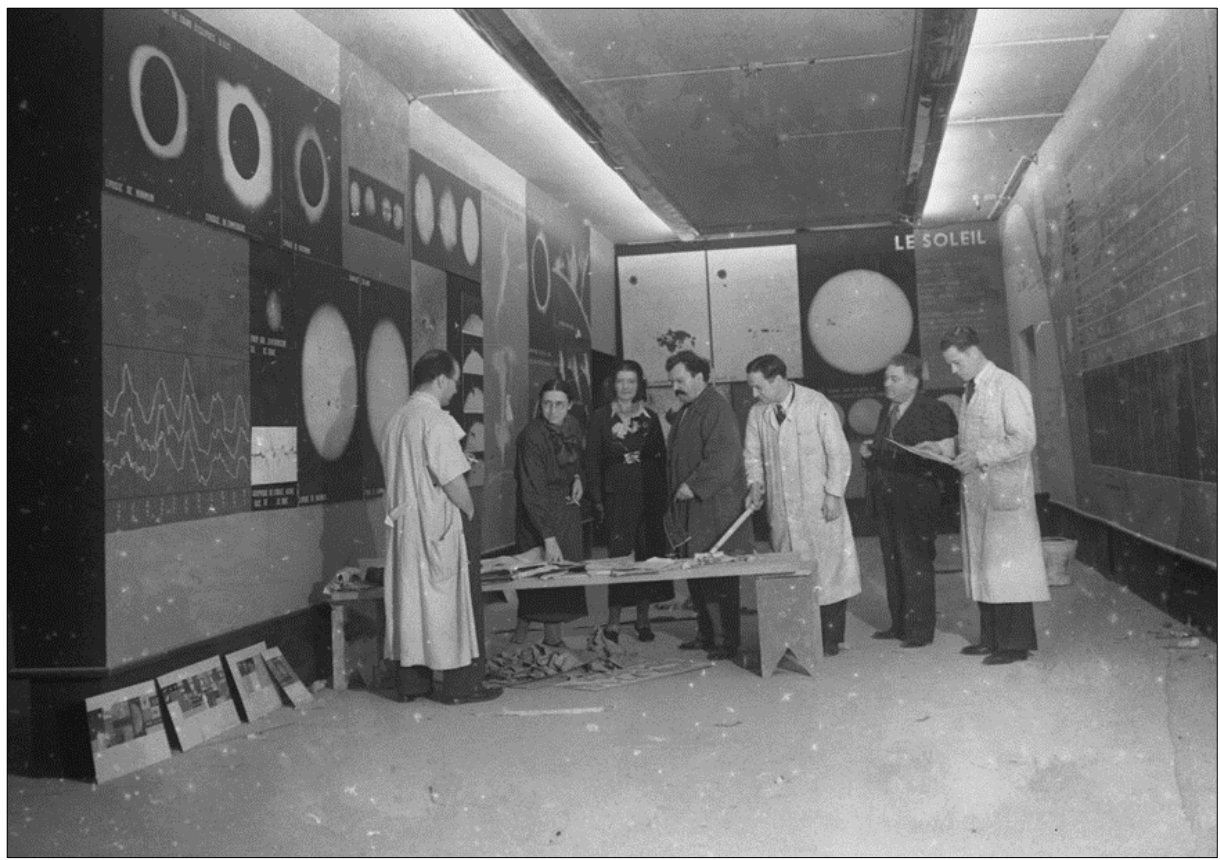

Figure 5: Reysa Bernson (second from left) visiting the solar exhibition that was being prepared by Paris Observatory staff at the Palais de la Découverte in 1937 (courtesy: Archives de l'Association Astronomique du Nord; © Association Jonckheere).

## 5 SCHOOL ACTIVITIES

In addition to the general public, Bernson also brought astronomical information to a great many schools (see Figure 3). At the time, astronomy did not form a large part of the school programs, and what they included was mostly unattractive and not in line with the latest discoveries—besides, it was often discarded through lack of time. Previous efforts to change the program, notably by SAF members, had failed. Trying another way, Bernson rather proposed one-shot events, in the form of a lively overview of contemporary astronomical knowledge of the Universe (Solar System – planets, stellar system – the Milky Way, galactic system – the expanding Universe), using lots of photographs. She started with students from secondary schools and students training to be teachers (i.e. young people 14–20 years old). She chose them for being 'an easy-to-handle audience' (in contrast to unruly children) and for the ease of gathering them in large numbers, to avoid repeating her talk to consecutive classes.

Aside from a basic goal of providing astronomical information, she was hoping to raise the bar by giving students a broader view. She wrote:

Doesn't all teaching consist in the study of human beings, whether individuals or groups, in their make-up and actions, and of the world in which they live? … Wouldn't few notions [give] the possibility of considering things and events from another angle, with another scale of measurement? And we believe that the sudden widening of the mental horizon that would result from this (and what a widening!) should prevent us from neglecting to introduce these notions into all teaching.

… give them a taste for a higher ideal, accustom them to broader horizons. (Bernson, 1937a).

In 1937, after a decade of experience, Bernson summarized her recommendations to do this kind of activity. First, the talk should be presented as a 'distraction', something not officially in the mandatory programs hence taking students away from school duties for a while.

This is a simple but effective psychological effect, but also a way to counter and use at one's advantage the non-inclusion of astronomy in the official program. Then she recommended

> a certain ease of elocution ... we think it's best to avoid the academic lecture, carefully drawn up ... have the broad outlines of your subject in your head and consult your notes as little as possible (not at all would be even better) ... don't look too stern or official. Giving it the form of a familiar, almost improvised talk, with the occasional anecdote to take the audience's mind off things for a moment, is a method that has always given us excellent results. Above all, you have to live your subject in front of the audience, not read it. (Bernson, 1937a).

After an initial trial with school classes made in 1927, she performed her school activities between 1929 and 1937. Although she mostly went to the schools themselves, doing the overview presentation in front of students, she also welcomed a few classes for observations at the AAN terrace observatory and even conducted a special science movie projection to 800 students. She kept a meticulous count of these activities (notably recapped in the annual summary of activities in the *AAN Bulletin*) so that she could assert having reached 12000 students in total through her secondary school presentations only (Bernson 1937b: Figure 3). These presentations were done in schools in the north-east quarter of France (Pas-de-Calais, Somme, Alsace, Lorraine, Burgundy, Paris area …). To those ones can be added a few presentations in Belgium although their exact number remains unknown since they do not appear in her counting statistics. It may be worth noting that she mentioned that she was not paid for all this work, but only got her travel expenses reimbursed.

While she had initially targeted only secondary school students, she decided in the end to try a few activities in primary schools, usually on request. They usually combined a short presentation and an observing session. To assess their efficiency, she added a feature for such cases: asking feedback reports. This allowed her to demonstrate that even young students could understand the basic astronomical features, comforting her in her broadening efforts. At least two cases are detailed in her articles, with excerpts of children's reports: one in Boulonnais schools in 1929 (Bernson, 1929b) and one in a school at the Belgium-France border in 1933 (Bernson, 1937a). The former case was in addition to her secondary school activities in the area, but she was amazed by the children's feedback—mentioning in passing that

> the girls gave infinitely fuller and more detailed accounts of these sessions than the boys, despite being younger in general. (Bernson, 1929b).

In the latter case, she received a request from a teacher after the republication of her 1929 experience in a specialized teacher magazine and the unexpected shootings stars of October 1933. Because of bad weather, however, she could only come in late November. That teacher was using 'active learning' (then called 'école nouvelle'), a pedagogical method that insisted that students be actors of their learning, something that she found particularly appealing as it was in line with her own tenets. She had already contacts with active learning people, but after this experience, she was invited for several teacher conferences in both France and Belgium.

Along with her philosophy to "… give as many people as possible the chance to discover the wonderful world around us …" (Bernson, 1937b), she tried a few original experiences, like sessions for blind students in 1932, for which she prepared hands-on features (Bernson, 1933c) and for sick/disabled young people in sanatoria in northern France in 1938 (Bernson, 1938a). Finally, one may also note her interventions on the French school radio channel, which were also much appreciated:

> We do not ask that the school radio educates while having fun when it is too difficult to achieve it, but we are happy to congratulate it when it succeeds in doing so. The Eiffel Tower station spoke about planets in a way to interest young and old alike. This session was conducted by Miss Reysa Bernson, the secretary general of the planetarium … [and] astronomy became charmingly crystal-clear. (Wahl, 1938).

## 6 SCOUT ACTIVITIES

Having targeted amateur astronomers, school students of various levels, and the public at large, Bernson tried to promote interest in astronomy to a last public group: scouts. Aside from the obvious additional possibility to reach young people, she viewed several advantages specific to scouting, an "… interesting method for educating the youth …" which recommends the "… study of nature … [with] constantly alert attention … [and] knowing how to use one own's eyes." (Bernson, 1939c). Her efforts were actually triggered by an article she read in *L'Astronomie* that related one case of astronomical out-reach for scouts (Richardot, 1932), but she went much further than this isolated experiment. She started by broadly disseminating plans for a cheap (25Frs), do-it-yourself refractor which she designed with René Réant, an AAN member (Bernson, 1933d). At the same time, she also invited the various local scout groups in the Lille area (Catholic, Protestant and lay) to observing sessions at the AAN Observatory.

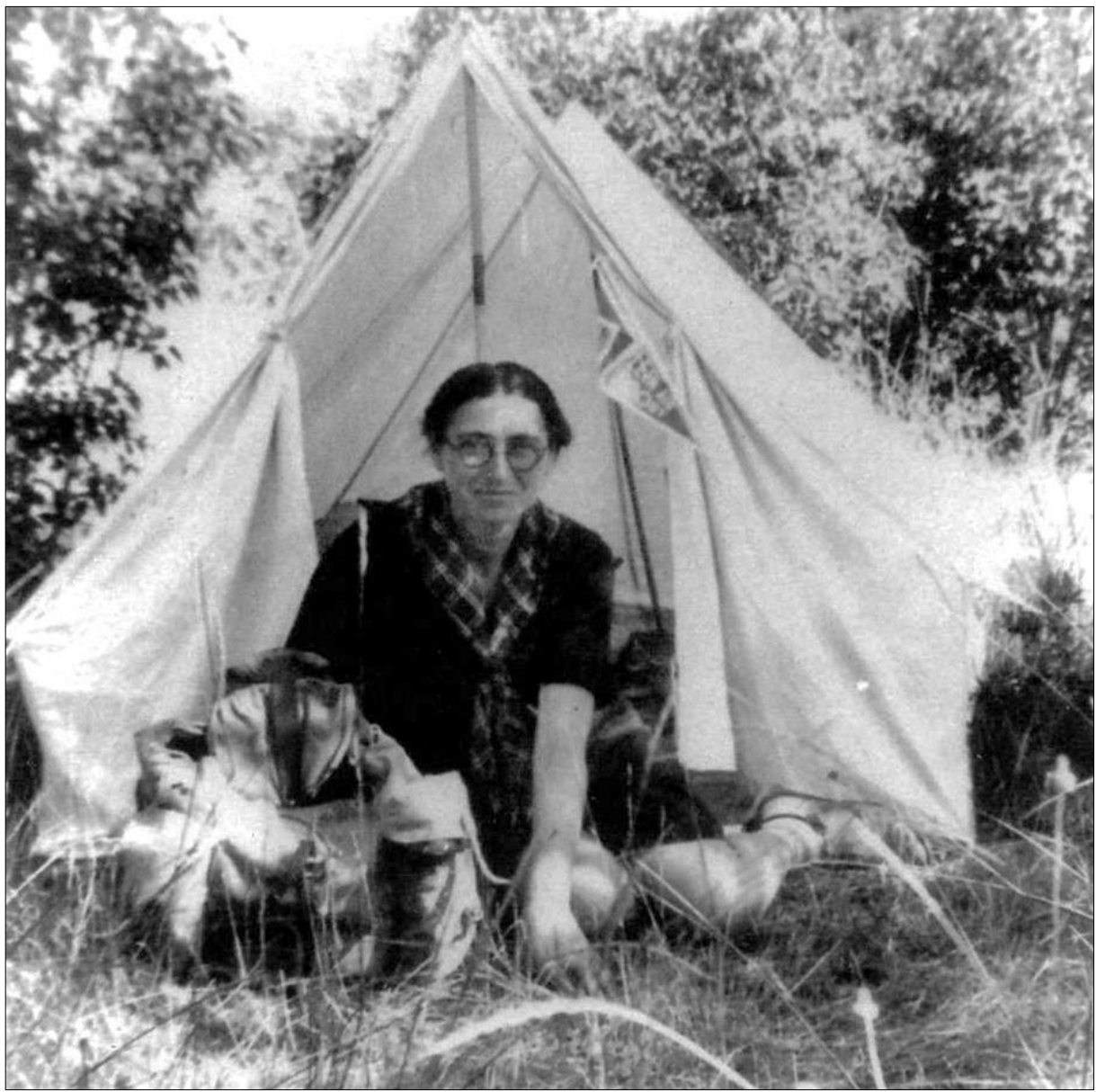

Figure 6: Reysa Bernson during a scouting activity (courtesy: Archives de l'Association Astronomique du Nord; © Association Jonckheere).

A few years later, she delved deeper into scout activities (e.g. see Figure 6). She led a reform of the 'badge' related to astronomy, proposing a clear program for it and a splitting from meteorological matters (for which another, separate, badge was then created). From 1937 to 1939, she also wrote[9] a series of articles in French scout magazines, but mostly *L'éclaireur de France*. Following the example of her childhood magazine *Les Petits Bonshommes*, she decided to repeat the use of a lively dialogue, although she exchanged the grandfather-and-grandson pair for a team of two young scouts, an experienced one and a more naïve one (called 'clever microbe'). In her articles, many themes are directly linked to scout activities (e.g. finding cardinal points and/or the time using the Sun, a compass and/or a watch), but she also presented basic facts on planets and stars. As in her childhood magazine, she proposed simple experiments that could be done by oneself, e.g. going around a colleague to see that he's projected in different directions as the Sun in its yearly course around the zodiac, lamp and balls to reproduce lunar phases, eclipses, or Venus appearing as a morning/evening star. She also used easily understandable comparesons, e.g. a white wall being invisible at night just as planets are when not illuminated by the Sun.

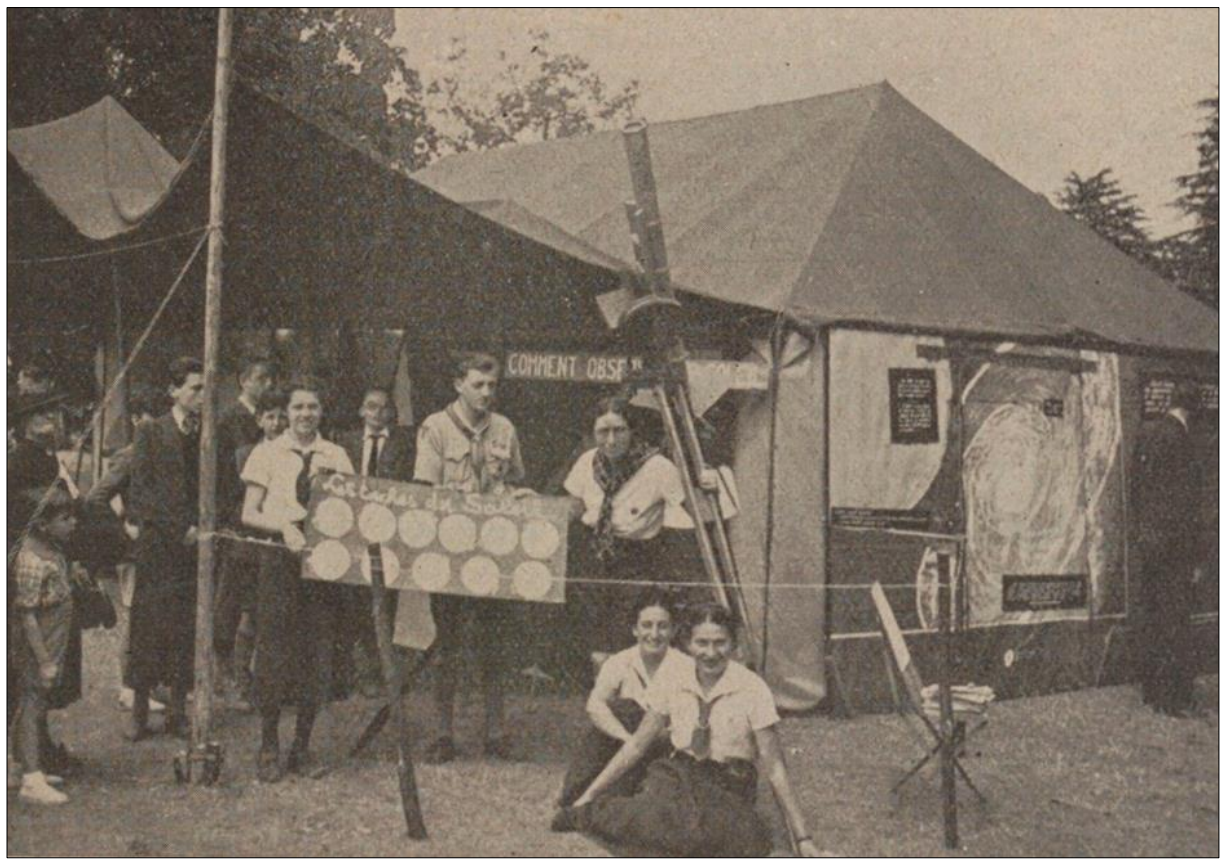

Figure 7: The Astronomical Center at the Scout Festival of June 1938 in Paris, with the young scouts from the Camille Flammarion group and their mentor, Reysa Bernson (after Bernson, 1938a: 413; courtesy: Gallica).

All the time, however, she actually taught the scientific method more than simple facts, insisting on a critical mind and one's own reasoning:

> Shouldn't a true scout always act intelligently, seeking to understand as much as possible what's going on around him and what he's doing himself? It's good for parrots to learn a recipe by heart, good for bandar-logs to mechanically repeat gestures they've seen done without knowing why they're doing them. You, microbe, are a scout, not a bandar-log, aren't you? So I'm not telling you to 'do this or that, and you'll find north'. I'm telling you how what you can see in the sky helps you find your way by doing this or that. And having understood that, I hope you won't stand there like you did before – because, even if you can't remember the exact recipe, you'll just have to think it through. (Bernson, 1939d).

> Great, buddy! Now that's what I call being a smart guy, a guy who doesn't just watch and record what he's seen without looking any further, but who asks himself why it's the way it is and not any other way. (Bernson, 1937c).

In particular, in her articles, she does not hesitate to let the naïve scout err at first in his reasoning. Then the experienced one asks questions until the naïve scout realizes by himself his errors, and then the experienced one provides a complete explanation.

In Spring 1938, the explorer Henri Lhote (1903–1991), then head of the 'Centre Naturalste des Éclaireurs de France', asked her to create a scout 'star team'. She accepted and the team was finally named the 'Camille Flammarion group' (Bernson, 1938a). She had meetings every two weeks with boy and girl scouts, to teach them the bases of astronomy. From the end of May, a core group had more frequent meetings with her, with the aim to present an 'astronomical center' at the Scout Festival of 26 June in the Parisian garden called 'Jardin d'Acclimation' (Figure 7). This was an impressive exhibition, with a repeated series of four talks using an armillary sphere and a scaled Solar System, observations of the Sun using a 108-mm refractor, a workshop to build small refractors, a stand selling elementary astronomy books, large sky maps and posters presenting the respective sizes of planets and stars as well as the location of the Earth in the Solar System and of the Sun in the Milky Way. Everything had been built and was animated by the boy and girl scouts of the Flammarion group. It was a huge success, and so they mounted the same exhibition at the Youth Festival in July of that same year.

Finally, at the end of 1938, she tried another experiment. In a two page article (Bernson, 1938b), she presented the lunar eclipse of 7 November, providing a list of tasks to be tried by the scout groups: measuring how the shadow advances with respect to lunar features (using a lunar map and a watch calibrated to the time announced by radio channels), noting the color and the visibility of lunar features in the shadow, and spotting stars in neighboring constellations over time (faint stars becoming visible when the Moon was fully occulted). This was a long list for children—she was not targeting the educated Flammarion group, this time—but she received many reports from all over France. Some of them included attempted photographs, which she had not requested, showing that some groups tried to go even further than requested, which she appreciated. She provided feedback, avoiding empty laudatory statements (Bernson, 1938c). Indeed, she congratulated the children for their efforts by pointing out specific achievements, but she simultaneously kept a critical eye, pointing out some errors in a constructive way (e.g. wrong time noted or missing information) to avoid repeating them next time. This experiment was also presented as a poster at the SAL inaugural day (Bernson, 1939b). This was a clear demonstration that even young, unexperienced observers could provide valuable information if correctly guided.

## 7  CONCLUDING REMARKS

Between the two World Wars, France saw an expansion of the public interest in astronomy, thanks to the increasingly numerous opportunities provided by amateur societies, media and museums. Reysa Bernson was one prominent figure during that epoch. She was involved in a broad range of activities, with the aim of disseminating astronomy to as many entities as possible: amateur astronomers, schools, specific groups such as people with disabilities or scouts, and the public at large. Using all possible means available at the time (direct observations, small experiments, radio talks, projections of movies or photographs, newspaper articles), she did it in a lively way, with simple yet precise and understandable information, not hesitating to add anecdotes or a humorous touch of her own. In all this, an obvious—yet maybe surprising considering the century that has elapsed since then—similarity with current outreach practices and philosophy can be drawn. Her expertise led her to head the first French planetarium in 1937 for the World Exhibition organized in Paris that year. Over a decade there is no doubt that she reached tens of thousands of people, an admirable achievement for that time, or even with current standards. In addition, she performed a lot of observations, notably contributing a large fraction of the variable star observations recorded by the AFOEV for female observers between 1920 and 1940. This notably led her to suggest the splitting of novae into two 'families' in 1935, simultaneously yet quite independently from similar proposals made by a few professional astronomers. Clearly, she was the most prolific female amateur astronomer of the time, as well as one of the few persons at the forefront of astronomy outreach in France between the two World Wars.

At the time her industrious work gained her some recognition by her peers. She occupied high positions in student organizations and enjoyed support from authorities, especially Albert Châtelet (1883–1960), the Dean of Lille University. That she was appreciated also clearly appears through texts published in many newspapers and magazines of the time. Additionally, she was the one invited to be the patron of the SAL, a new amateur astronomical society created in Belgium in 1938. Finally, she received several SAF awards: the Henri Rey Prize in 1932 and the de la Guette Observatory Prize in 1937, as well as the Commemorative Medal (this one jointly with her planetarium team). It may be noted that amongst the 187 SAF prizes awarded between 1897 and 1945 (Anonymous, 1945), only 10 went to individual women and she is the only one who got two. Such a repeated reward was as infrequent for men, as only a dozen of them received several SAF awards over the same period of time. In her planetarium team, this was the case for André Hamon and Auguste Budry; apart from them, it is notably the case for Lucien Rudaux and Ferdinand Quénisset.

In parallel, she had to face several setbacks, such as the halting of her radio series in 1933, the erasing of her name from eclipse expedition reports in 1936, and the total disappearance of her name from AAN activities and *Bulletins* after 1936. The exact cause of these problems was never clarified but jealousy for all the public appreciation and recognition she received (an enduring feature suffered by astron-omers doing outreach) as well as gender and/or racial prejudice—which were rather common at that time—may have played a role. The largest injustice was of course her tragic and premature death. Because her mother was Jewish, Reysa Bernson started to hide from the Germans in 1940, but she was finally arrested in 1944 along with her mother. They were deported to Auschwitz extermination camp in convoy #69 from the Drancy camp, being most probably eliminated upon arrival (Delmaire and Faidit, 2017). In any case, there was no news of her after the War (Anonymous, 1949; Flammarion, 1946).

After World War II, Reysa Bernson's name disappeared from common use, with only a few scattered mentions (mostly in Belgium, and notably at anniversaries of the foundation of the SAL). In recent years, however, her work has been rediscovered, and a few French publications mentioning it have started to appear (e.g. Barré-Lemaire, 2019; Delmaire and Faidit, 2017; Faidit, 2015; 2019; Mathieu, 2022). There were also two important naming efforts in 2018: a small alley in Lille (Conseil Municipal, 2018) and asteroid #21114, an asteroid of the Flora family, discovered by the Belgian astronomer E.W. Elst (1936–2022) in 1992 (MPC, 2018). Bernson's name was also included in a website listing 100 female scouts (https://astrales.fr) and in the 2021 calendar of the associated Scout organization (Mathieu, 2022). She probably deserves more, and it would only be justice, for example, to name a French planetarium after her (notably the one at the Palais de la Découverte in Paris).

## 8  NOTES

1. Her birth date is often quoted as 28 September 1904, but the Lille municipal archives actually show that this is the date at which the father came to the Town Hall to register the baby, who was born two days before, on the 26th – see https://archives.lille.fr/ark:/74900/dr91fgp4hbkw/5c32c648-2096-4d24-ae2c-d8b86bfbc899
2. The sole exception to this rule is the list of victims kept at Yad Vashem Memorial (https://deportation.yadvashem.org/?language=fr&itemId=5092640).
3. The date is not mentioned in the article, but this was the only solar eclipse visible from Lille during her childhood.
4. It is not known why Bernson did not do the talk herself, as in the previous year, but maybe this was a political move, showing that the SAF backed the idea as Mrs Flammarion was held in high regard by the Society.
5. This might even have saved her life, as professional Jewish astronomers received help to hide from German round-ups and arrests during World War II. One example is Marie Bloch, the previously mentioned Assistant at Lyon Observatory who was only two years older than Reysa Bernson and who continued a fruitful career after the War.
6. This monthly chronicle was done on request, from December 1932 to June 1933. It was halted despite its success "… at the request of the management of the radio station PTT Nord." The series resumed later, but with a new speaker "… at the express wish of the directors of the radio association …" (Comité de l'AAN, 1933). The reason for stopping and not retaining a successful speaker remains undocumented.
7. By way of comparison, we may point out that Mrs Gabrielle Flammarion, the widow of the famous popularizer, was very much respected in the astronomical circles of the 1930s. She often organized or intervened in meetings and was given numerous opportunities to speak (and she *was* listened to). She was identified as one of the few SAF heads, but she was never given the honour of serving as President of the Society (in fact, until now all Presidents have been male).
8. Delsemme and Oriano both became professional astronomers after World War II, specializing in comets and galaxies respectively. Nowadays, Oriano is better known as de Vaucouleurs. It is interesting to note that after 1937 and his meeting with Bernson he went and switched his family name to his mother's name, just as she had done. Both Delsemme and de Vaucouleurs have left moving accounts of their memories of Bernson, underlining their high respect for her (Delsemme, 1989; Lightman, 1988).
9. These articles are available on this site: https://astrales.fr/reysa-bernsonastronome-passionnee-de-transmission-aux-jeunes/ - Note that they were always signed 'asteroid #1021' (i.e. the asteroid named after Camille Flammarion). However, the table of content of *Le Chef*, Volume 196 (April 1939), lists her name as the real author of the articles signed 'asteroid #1021', so the identification is not in doubt.

## 9  ACKNOWLEDGMENTS

The author warmly thanks André Amossé and the amateurs taking care of Lille Observatory, Stephanie Laden, Nathalie Barré-Lemaire and Georges Wlodarczac of Lille University, Olivier

Borsus and Alain Detal from Liège University, Jean Minois and Dominique Naillon from the AFOEV, Jean-Claude Bercu from the SAF, Florian Mathieu, Jacqueline Eidelman and Andrée Bergeron (CNRS), and Myron A. Smith. She acknowledges the use of AFOEV, ADS, Gallica, Roubaix mediathèque, archive.org, and Jstor databases. Note that all translations were made by the author. Note that most of the entries in References section can be found on ADS (https://ui.adsabs.harvard.edu/) and/or Gallica (https://gallica.bnf.fr/).

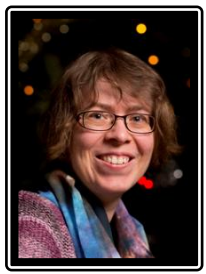

**Dr Yaël Nazé** is a FNRS Senior Researcher working at the University of Liège in Belgium. She earned her PhD in 2004 and became an IAU member in 2012.

Most of her research concerns the observational properties of massive stars, but she also carries out some multidisciplinary research on historical or sociological aspects of astronomy. In this context, she has notably published on Newton's chronology and astronomical knowledge in the Belgian public, Max Ernst's work on Wilhelm Tempel, and eighteenth century sci-fi novels by female writers.

Her most recent books are *Art & Astronomie* (2015) and *Astronomie de l'Étrange* (2021). A new book will come out in 2024.